\documentclass[]{aastex}

\usepackage{emulateapj5}
\usepackage{apjfonts}

\newcommand{\mpcc}{{\rm Mpc^{-3}}}

\begin{document}

\title{
The Luminosity Function of {\sl IRAS} PSC$z$ Galaxies
}

\author{
Tsutomu T. Takeuchi\altaffilmark{1,2}, 
Kohji Yoshikawa\altaffilmark{3}, and
Takako T. Ishii\altaffilmark{4}
}

\altaffiltext{1}{National Astronomical Observatory of Japan,
Mitaka, Tokyo 181--8588, JAPAN; takeuchi@optik.mtk.nao.ac.jp.
}

\altaffiltext{2}{
  Research Fellow of the Japan Society for the Promotion of Science
}

\altaffiltext{3}{Research Center for the Early Universe (RESCEU),
The University of Tokyo, Bunkyo-ku, Tokyo 113--0033, JAPAN;
kohji@utap.phys.s.u-tokyo.ac.jp.
}

\altaffiltext{4}{
  Kwasan and Hida Observatories, Kyoto University, 
  Yamashina-ku, Kyoto 607--8471, JAPAN; ishii@kwasan.kyoto-u.ac.jp.
}

\begin{abstract}
We estimated the luminosity function (LF) of {\sl IRAS} galaxies in the PSC$z$
catalogue.
The faint end of the PSC$z$ LF is slightly steeper than
that of the LF derived by Saunders et al.\ (1990; S90).
Using an analytical form for the LF used by S90, we obtain
the following parameters: 
$\alpha = 1.23 \pm 0.04$, 
$L_*=(8.85 \pm 1.75) \times 10^8 h^{-2}\;L_\odot$, 
$\sigma =0.724 \pm 0.010$, and
$\phi_* = (2.34 \pm 0.30) \times 10^{-2}h^3\;\mpcc$.
We also examined the evolution in the sample by a simple assumption
$\phi_*(z) \propto (1+z)^P$, and found $P=3.40 \pm 0.70$.
It does not affect the three parameters, $\alpha$, $L_*$, and $\sigma$,
but $\phi_*(z=0)$ is overestimated up to $\sim 15$~\% if we ignore evolution.
We estimated the temperature dependence of the LF.
The LFs of warm and cool galaxies are quite different: 
the LF of warm galaxies has a very steep faint end with $\alpha =1.37$.
We also discuss a lump found at the brightest end of the LF.
\end{abstract}

\section{INTRODUCTION}

Galaxy luminosity function (LF) is an important tool to characterize 
the statistical properties of galaxies.
The far-infrared (FIR) LF enables us to evaluate the properties of 
interstellar dust and its heating sources in various galaxy populations.
The Infrared Astronomical Satellite ({\sl IRAS}) has provided us a uniform
FIR all-sky survey of local galaxies.
In addition to its own importance, the FIR LF has also come to the limelight 
in conjunction with the evolution of FIR galaxies.
Evaluation of their evolution plays a crucial role in understanding the 
cosmic star formation history hidden by dust (e.g., 
\citealt{franceschini01,takeuchi01a,takeuchi01b,rowan01,granato00,totani02,takeuchi02}).

In spite of many attempts to estimate the LF of {\sl IRAS} galaxies 
(e.g., \citealt{lawrence86,soifer87}; \citealt{saunders90} (S90); 
\citealt{isobe92}), there is yet room for improvement in its exact shape, 
especially at the faint end.
The best way to overcome this difficulty is, of course, a statistical analysis 
of a huge homogeneous dataset.

The PSC$z$ catalogue \citep{saunders00} (S00) is the largest well controlled 
redshift sample of {\sl IRAS} galaxies to date.
The catalogue is complete to 0.6~Jy at $60\;\mu$m and contains 15411
{\sl IRAS} galaxies covering 84~\% of the sky.
This is an ideal dataset to estimate the exact shape of the LF of {\sl IRAS} 
galaxies.

In this {\sl Letter}, we estimate the LF from the PSC$z$ catalogue
using the statistical package for LF estimation developed by us 
(\citealt{takeuchi00}; hereafter T00) and examine the exact shape of 
the FIR LF.
We use the cosmological parameter set $(h, \Omega_0, \lambda_0) = 
(0.7, 0.3, 0.7)$ except otherwise stated.

\section{ANALYSIS}\label{sec:analysis}

\subsection{Data Description}

\subsubsection{Flux density}

The data provided by S00 include (1) PSC flux, (2) point-source-filtered 
{\sc addscan} flux, and (3) extended (coadded or extended {\sc addscan}) 
flux for all the sources.
First we should note that the flux densities of nearby extended sources 
can be seriously underestimated because of the fixed aperture of {\sl IRAS}.
Indeed, Figure~2 of S00 shows a clear trend of underestimation in PSC flux or 
point-source-filtered flux.
We used (3), coadded or extended {\sc addscan} flux, that
has been carefully measured and homogenized by S00.

\subsubsection{Local Group galaxies}

The main catalog of S00 explicitly excludes the Local Group (LG) galaxies,
and they are separately compiled in another catalog `{\sf ilg.dat}'.
There is a possibility that the exclusion of these galaxies 
slightly affects the faintest-end slope estimate of the LF, because
they generally have low luminosities.
In order to include the LG galaxies we adopted the distances presented by
\cite{vandenbergh00}, adjusted for the cosmological parameters used.
We compared the LFs estimated from the datasets with and without LG galaxies,
and obtained completely the same results:
their inclusion caused no difference at all.
For Virgo infall correction, we adopted a flow model very similar to 
model V3 of S90.

\subsection{$K$-correction}

{}To make $K$-correction, we used extended fluxes of $S_{25}$, $S_{60}$, 
and $S_{100}$ tabulated in the PSC$z$ catalogue and fitted a 2nd-order
polynomial to the flux densities for each galaxy:
\begin{eqnarray}
  \log \tilde{S}_{\lambda} (\log \lambda) &=& 
    \sum_{i=0}^{2} a_i (\log \lambda)^i \;,\\
  \log S_{60}^{\rm em} &=& \log \tilde{S}_{[60(1+z)]}
\end{eqnarray}
where $\tilde{S}_{\lambda}$ is the interpolated flux density at 
wavelength $\lambda \, \mu$m, superscript `em' means that the flux is 
measured at the emitted wavelength, and coefficients $a_i$ are
estimated from the observed fluxes $S_{25}$, $S_{60}$, and $S_{100}$.

\subsection{Statistical Methods}

T00 extensively discussed four nonparametric statistical methods 
for estimating galaxy LF.
The methods are 1) $1/V_{\rm max}$ method \citep{schmidt68,eales93}, 
2) \cite{efstathiou88} (EEP) method, 3) \cite{choloniewski86} method, 
and 4) improved $C^-$ method \citep{lyndenbell71,choloniewski87}.
The associated error is estimated by a bootstrap resampling (see T00).
Equation~(\ref{eq:lf}) has been verified to be appropriate for a FIR LF 
(see Appendix~D of S90).
Hence, in Section~\ref{sec:lf} we also show the parametric fit for the LF 
by the parameter estimation advocated by \cite{sandage79}, using 
\begin{eqnarray}\label{eq:lf}
  \phi (L) = \phi_* \left( \frac{L}{L_*} \right)^{1-\alpha}
  \exp \left[ -\frac{1}{2\sigma^2} 
    \left\{\log \left(1+\frac{L}{L_*}\right)\right\}^2\right]\;.
\end{eqnarray}
We note that, while the $C^-$, EEP, and the Cho{\l}oniewski methods are
robust against density fluctuation, the $1/V_{\rm max}$ method is not 
accurate if a density inhomogeneity exists in the sample, as pointed out 
by previous studies.

\section{THE PSC$z$ LUMINOSITY FUNCTION}\label{sec:lf}

\vspace{0.5cm}
\centerline{{\vbox{\epsfxsize=8.5cm\epsfbox{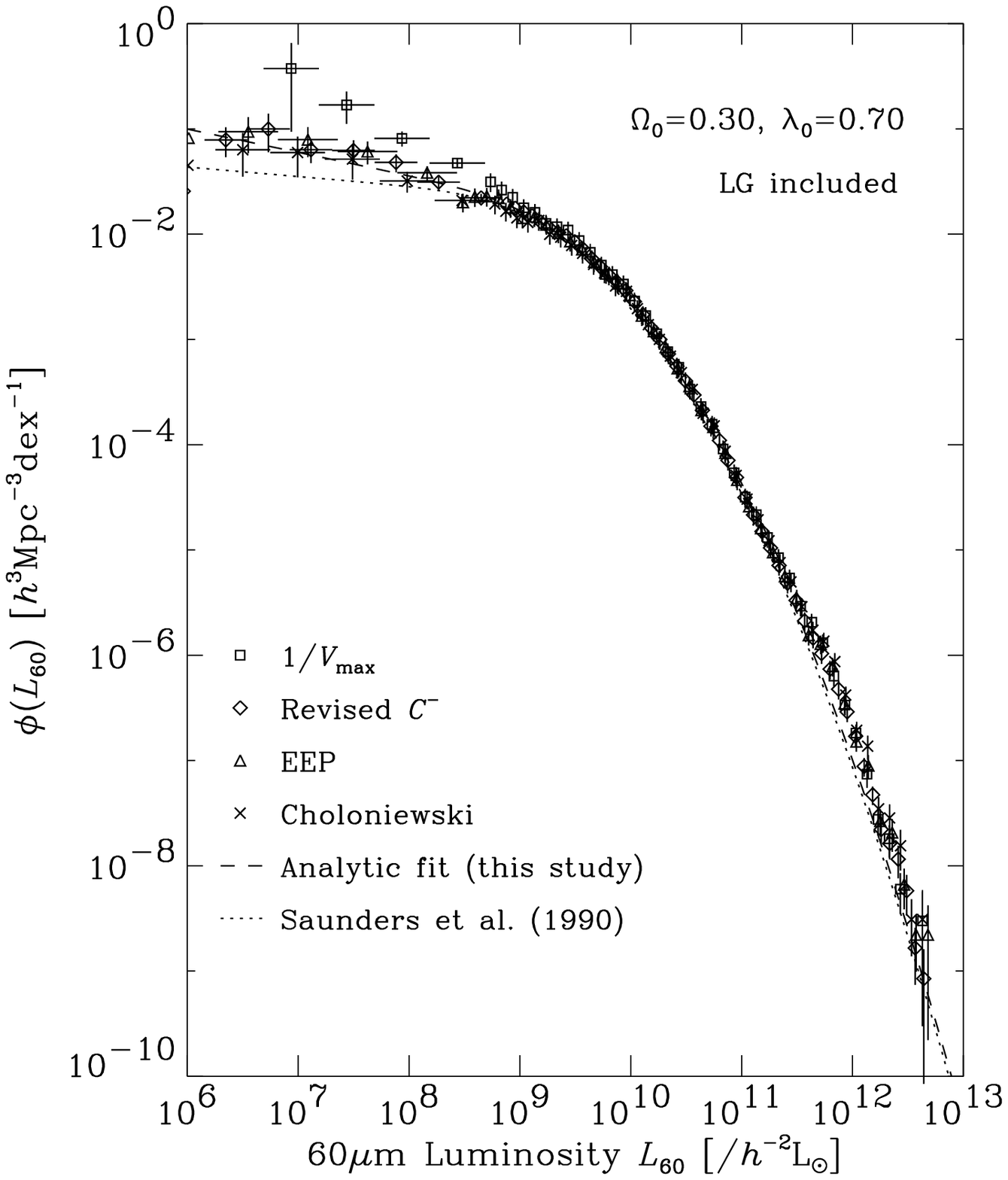}}}}
  \figcaption{
  The luminosity function (LF) of {\sl IRAS} PSC$z$ 
  galaxies.
  The squares, diamonds, triangles, and crosses represent the results obtained 
  with $1/V_{\rm max}$, improved $C^-$, EEP, and Cho{\l}oniewski methods, 
  respectively.
  The faint-end estimates ($10^{6}\mbox{--}10^{9}\;L_\odot$) are rebinned
  to reduce statistical fluctuation.
  The dashed line describes the analytic LF derived in this work.
  The parametric result of \cite{saunders90} is also shown by
  the dotted line.
\label{fig:lf}}
\vspace{0.5cm}

Figure~\ref{fig:lf} is the FIR LF of the PSC$z$ catalogue.
Different symbols represent the results obtained by different estimators.
The dashed line describes the analytic LF derived in this work.
The parametric solution of S90 is also shown in Figure~\ref{fig:lf} by the 
dotted line.
We see that the estimates by the three inhomogeneity-insensitive methods 
show excellent agreement with one another,
and at a faint regime $L_{60} \la 10^{9}\;L_\odot$, 
$1/V_{\rm max}$ estimate significantly deviates upward because of 
the local density enhancement.
We note that LG galaxies are included in the sample to produce 
Figure~\ref{fig:lf}.

S90 proposed the following values for Equation~(\ref{eq:lf}):
$\alpha = 1.09 \pm 0.120$, 
$L_* = (2.95^{+2.06}_{-1.21}) \times 10^{8}h^{-2}\;L_\odot$, 
$\sigma = 0.724 \pm 0.031$, and
$\phi_* = (2.6 \pm 0.8) \times 10^{-2}h^3\; \mpcc$.
These values were obtained from the compiled redshifts of 2818 {\sl IRAS} 
galaxies.
We find that the bright end of the PSC$z$ LF is remarkably well described 
by the analytic form estimated by S90.
The normalization also nicely agrees with that of S90.
At this stage we ignore the possible effect of galaxy evolution, which 
we will discuss in Section~\ref{sec:evolution}.

However, the faint end deviates from the result of S90.
The three inhomogeneity-insensitive estimates show consistently the same trend.
Hence the steep faint end of the PSC$z$ LF is not caused by density 
enhancement, and we conclude that it is an intrinsic statistical property 
of the local {\sl IRAS} galaxies.

We also estimated parameters for the analytic fit expressed by 
Equation~(\ref{eq:lf}) as follows:
\begin{eqnarray}
  \begin{array}{lclcll}
    \alpha &=& 1.23  &\pm& 0.04\,,                  &{}          \\
    L_*    &=& (8.85 &\pm& 1.75)\times 10^8 h^{-2}  &[L_\odot]\,,\\
    \sigma &=& 0.724  &\pm& 0.01\,,                 &{}          \\   
    \phi_* &=& (2.60 &\pm& 0.30)\times 10^{-2}h^{3} &[\mpcc]\;.
    \end{array}
\end{eqnarray}
The uncertainty for each parameter is estimated from the relative 
marginalized logarithmic likelihood.
The log-likelihood around the maximum likelihood solution asymptotically 
behaves like a Gaussian, hence we can estimate 1-$\sigma$ 
error from $\Delta \ln {\cal L} \equiv \ln {\cal L} - 
\ln {\cal L}_{\rm max} = -0.5$.
We note that $\alpha$ is larger than that of S90 (3-sigma significant), 
though within their quoted error, it is consistent with the estimate of S90.
The large $\alpha$ quantitatively describes the above-mentioned steep 
faint end.
On the other hand, we obtain almost the same value for $\sigma$, which
characterizes the bright-end shape, and $\phi$, which determines the 
density normalization.

\section{THE LUMINOSITY FUNCTION OF WARM AND COOL {\sl IRAS} GALAXIES}
\label{sec:tlf}

In this section we derive the LFs of relatively warm and cool galaxies in
the PSC$z$ sample of galaxies.
We chose the boundary of these classes to be $\beta = S_{100}/S_{60}= 2.1$, 
which is almost the median value for the sample.
\cite{helou86} has discussed the color--color diagram of normal {\sl IRAS} 
galaxies. 
{}From his Figure~1, we find that $\beta = 2.1$ is also equal to the median of 
his sample.
If we use a modified blackbody with emissivity index 1.5, $\beta = 2.1$ 
corresponds to the dust temperature $\sim 25$~K.
We refer to the galaxies with $\beta \le 2.1$ as `warm galaxies', 
and those with $\beta > 2.1$ as `cool galaxies' in this study.

S00 have assigned extended {\sc addscan} fluxes to all galaxies, 
we simply used these values to estimate $\beta$.
However, warmer galaxies tend to have larger luminosities 
(e.g., \citealt{takeuchi01a}), this procedure might introduce a complex
selection bias.
We reconstruct the complete subsample of each category by adopting 
a shallower flux limit of $S_{60} \ge 0.912$~Jy.
The final sizes of the two samples are 5554 (warm) and 3421 (cool).

\vspace{0.5cm}
  \centerline{{\vbox{\epsfxsize=8.5cm\epsfbox{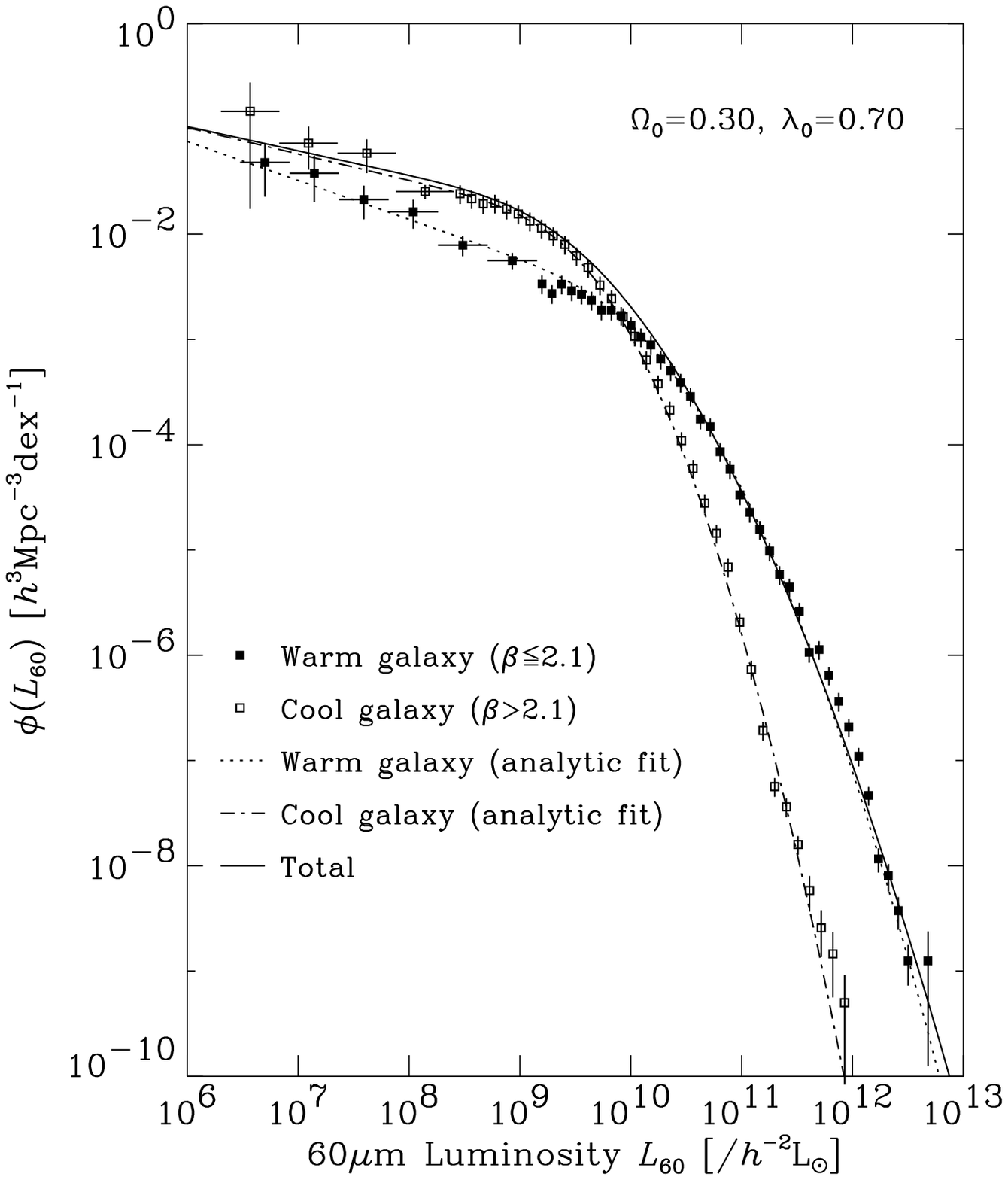}}}}
  \figcaption{
  The luminosity function (LF) of warm ($\beta \le 2.1$) and cool 
  ($\beta > 2.1$) galaxies.
  The solid curve describes the analytic expression of the LF of the whole 
  PSC$z$ galaxies.
  Filled and open squares represent the LFs of warm and cool galaxies, 
  respectively.
  Dotted and dot-dashed lines depict the analytic fits of the warm and cool
  galaxies.
\label{fig:tlf}}
\vspace{0.5cm}

The LFs of the warm and cool PSC$z$ galaxies are shown in Figure~\ref{fig:tlf}.
In Figure~\ref{fig:tlf}, we show only the estimates by EEP method for clarity;
but the other estimates show a good agreement with one anther.
The parameters of the analytic fit for warm galaxies are
\begin{eqnarray}
  \begin{array}{lclcll}
    \alpha &=& 1.37  &\pm& 0.05\,,                  &{}          \\
    L_*    &=& (5.10 &\pm& 0.90)\times 10^9 h^{-2}  &[L_\odot]\,,\\
    \sigma &=& 0.625  &\pm& 0.015\,,                &{}          \\ 
    \phi_* &=& (4.20 &\pm& 0.85)\times 10^{-3}h^{3} &[\mpcc]\,,
    \end{array}
\end{eqnarray}
and those for cool galaxies are
\begin{eqnarray}
  \begin{array}{lclcll}
    \alpha &=& 1.25  &\pm& 0.06\,,                  &{}          \\
    L_*    &=& (1.95 &\pm& 0.40)\times 10^8 h^{-2}  &[L_\odot]\,,\\
    \sigma &=& 0.500  &\pm& 0.020\,,                &{}          \\
    \phi_* &=& (1.85 &\pm& 0.37)\times 10^{-2}h^{3} &[\mpcc]\,.
    \end{array}
\end{eqnarray}
We also show the analytic fits for these two groups in Figure~\ref{fig:tlf}.
S90 performed a similar analysis of warm and cool galaxies. 
They first fit modified blackbody spectra to {\sl IRAS} flux data and 
defined dust temperature, and then divide the sample into galaxies with 
$T \ge 36$~K and $T < 36$~K.
Though they give systematically higher temperature than ours, their LF of 
each class is quite similar to ours; therefore, S90 and we probably treat 
nearly the same populations.

It is interesting to note that the faintest end of the warm galaxies 
increases steeply.
This means that a significant fraction of galaxies at luminosities
$L_{60} \la 10^7\; L_\odot$ can be low-luminosity but hot galaxies.
Because of their low luminosity, detailed case studies of their spectral 
energy distributions and next generation large FIR surveys (e.g., ASTRO-F,  
{\sl SIRTF}, and {\sl Herschel}) are important to reveal their physical 
properties.

\vspace{0.5cm}
  \centerline{{\vbox{\epsfxsize=8.5cm\epsfbox{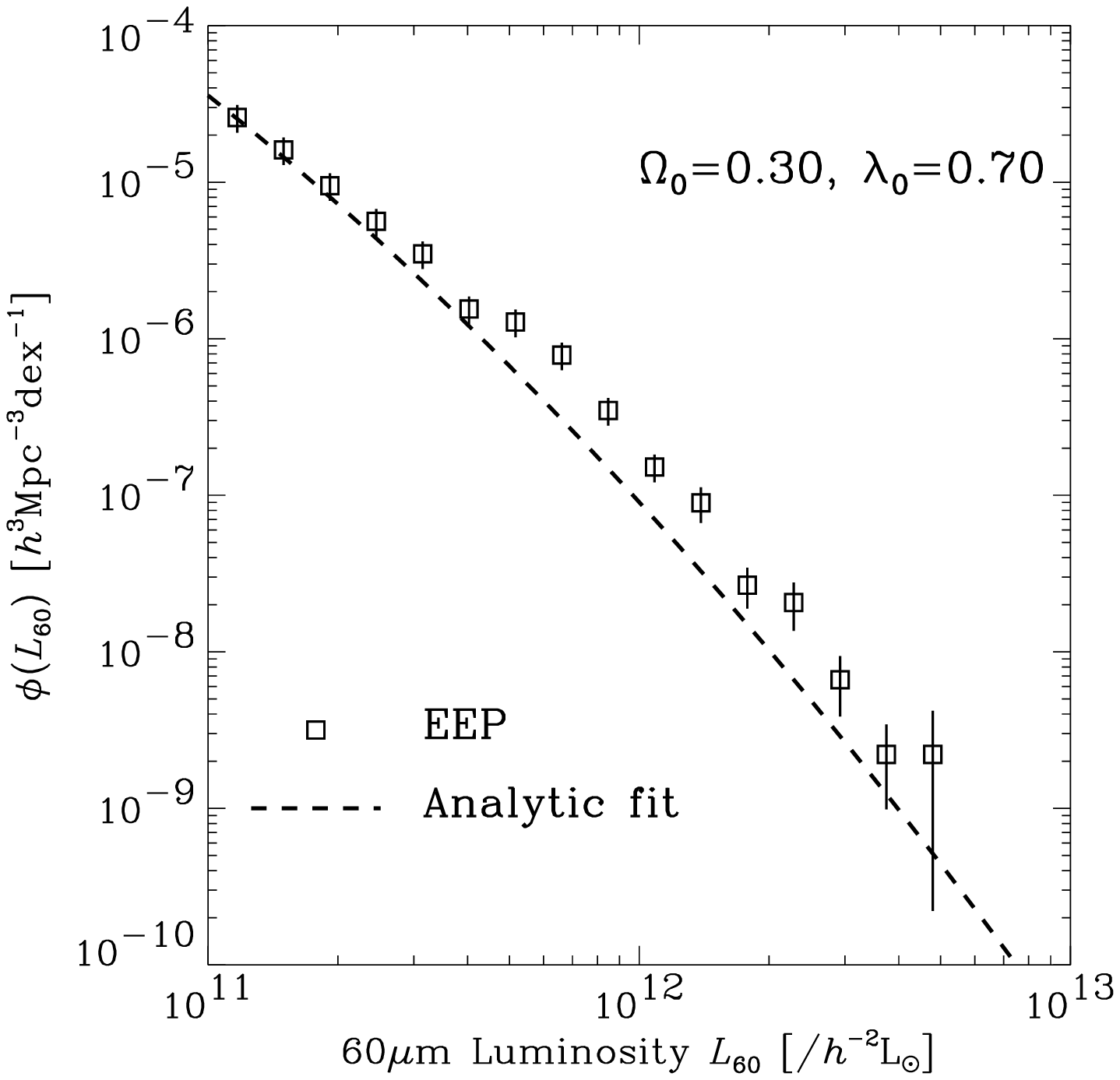}}}}
  \figcaption[f3.eps]{
  The zoom-up of the bright-end of the luminosity function.
  We only plot EEP estimates for clarity.
  A significant lump is found at luminosity $L_{60} \ga 10^{11.5}\; L_\odot$.
\label{fig:bright_end}}
\vspace{0.5cm}

\section{DISCUSSION}\label{sec:discussion}

\subsection{Possible Systematic Uncertainties in LF Parameters}

\subsubsection{Hubble constant}

We further consider some subtle effects on the parameter estimation. 
The effect of the choice of cosmological parameters is examined in S90.
Their Figure~5 illustrates the dependence of the estimated LF.
We should be cautious that the faint end slope is slightly affected by the 
choice of the Hubble constant, $h$.
But according to their Figure~5, if we use larger $h$, the estimated $\alpha$ 
will be smaller.
Hence, our steep faint-end slope is not due to this effect.

\subsubsection{Evolution}\label{sec:evolution}

\cite{rowan01} mentions the possibility that the parameters are 
affected by galaxy evolution of the sample.
Hence we examined this issue by dividing the whole sample into 
$z < 0.02$ and $z > 0.02$ groups.
We derived the LF of each group and found that the faint end of the LF of 
the total PSC$z$ sample is dominated by galaxies at $z < 0.02$.
The flux density limit of 0.6~Jy corresponds to $L_{60} \sim 10^{9.6}\;
L_\odot$ at $z = 0.02$ ($h = 0.7$), therefore the evolution toward higher 
redshift can hardly affect the estimation of $\alpha$.

On the other hand, evolution may affect the estimation of the 
normalization of the LF.
We estimated the evolution strength under the assumption 
$n(z) \propto n_0(1+z)^P$, where $n_0$ and $n(z)$ is the comoving 
number density of galaxies at redshift $0$ and $z$, respectively (S90).
If we assume that the LF is separable for $L$ and $z$ as 
$\phi(L,z) = n(z)p(L)$, then the likelihood is expressed as
\begin{eqnarray}\label{eq:likelihood_equation}
  {\cal L}(P|\{(L_i,z_i)\}_{i=1,\cdots,N}) &=& 
    \prod_{i=1}^{N} \frac{n(z_i)p(L_i)}{\int n(z)p(L_i)(dV/dz)dz} \nonumber \\
    &=& \prod_{i=1}^{N} \frac{(1+z_i)^P}{
      \int^{z_{{\rm max},i}}_{0} (1+z)^P(dV/dz)dz}\;,
\end{eqnarray}
where $p(L)$ denotes the probability density for a galaxy having 
a luminosity $L$, and $z_{{\rm max},i}$ represents the maximum redshift
to which a galaxy $i$ can be detected within the flux limit $S_{\rm lim}$.
We can estimate the strength of density evolution, $P$, by maximizing the 
above likelihood equation [Equation~(\ref{eq:likelihood_equation})].

We obtained $P=3.40\pm 0.70$.
This value is much smaller than that estimated by S90, but consistent with
the estimate of \cite{springel98}.
By using this value, we found that the density normalization $\phi_*(z=0)$ is
overestimated up to $\sim 10\mbox{--}15$~\% if evolution is ignored.
This systematic error is comparable to the statistical uncertainty of 
$\sim 12$~\%.
Thus, the density normalization parameter is $\phi_*=2.34$ if we properly
take into account the evolution.

\subsection{Bright End Population}

It is worthwhile to note the lump found in the bright end of the LF
in Figure~\ref{fig:lf} at $L_{60} \ga 10^{11.5}\;L_\odot$.
The zoom-up of the bright-end is shown in Figure~\ref{fig:bright_end}.
We only plot EEP estimates for clarity.
We consider which population of galaxies dominates the high-luminosity lump.
\cite{machalski00} presented the radio (1.4~GHz) LF from UGC \citep{nilson73}
and LCRS \citep{shectman96} galaxy samples.
Their radio LF clearly shows that it consists of two distinct populations:
star-forming galaxies and AGNs.
At $L_{1.4\,{\rm GHz}} > 10^{23.4} \;[\mbox{WHz}^{-1}]$ AGN contribution 
dominates the LF.
We can convert the radio luminosity to FIR one by the well-known radio-FIR
correlation \citep{condon92} and obtain the corresponding $L_{\rm FIR} 
\simeq 5.0 \times 10^{11}\;L_\odot$.
It agrees with the 60-$\mu$m luminosity at which the lump begins to appear.
So we suggest that the AGN contribution is attributed to the lump
of the LF at $L_{60} \ga 10^{11.5}\;L_\odot$.
Different energetics may be the cause of discontinuity in the LF shape.
It is interesting that both warm and cool galaxies show departure from 
the analytic fit at the same $L_{60}$ (see Figure~\ref{fig:tlf}).

\acknowledgments
We thank two referees for thoughtful comments and scrutiny that
have improved the quality of this paper very much.
We also thank H. Hirashita, T. N. Rengarajan, and M. Imanishi 
who gave us useful suggestions.
TTT has been supported by JSPS.

\end{document}